\documentclass[
reprint,
aps,
superscriptaddress,
pra,
]{revtex4-1}
\usepackage[utf8]{inputenc}
\usepackage{graphicx}
\usepackage{dcolumn}
\usepackage{bm}
\usepackage{ulem}
\usepackage{amssymb}
\usepackage{braket}
\usepackage{verbatim}
\usepackage{mathtools}
\usepackage{bbold}
\usepackage{cancel}
\usepackage[dvipsnames]{xcolor}

\begin{document}
\title{Optimization of Pulses with Low Bandwidth for Improved Excitation of Multiple-Quantum Coherences in NMR of Quadrupolar Nuclei}
\author{Jens Jakob S{\o}rensen}
\affiliation{Department of Physics and Astronomy, Aarhus University, Ny
	Munkegade 120, 8000 Aarhus C, Denmark}
\author{Jacob S{\o}gaard Nyemann}
\affiliation{Department of Physics and Astronomy, Aarhus University, Ny
	Munkegade 120, 8000 Aarhus C, Denmark}
\affiliation{Interdisciplinary Nanoscience Center and Department of Chemistry, Aarhus University, Gustav Wieds Vej 14, 8000 Aarhus C, Denmark}
\author{Felix Motzoi}
\affiliation{Department of Physics and Astronomy, Aarhus University, Ny
	Munkegade 120, 8000 Aarhus C, Denmark}
\affiliation{Forschungszentrum J\"ulich, Institute of Quantum Control (PGI-8), D-52425 J\"ulich, Germany}
\author{Jacob Sherson}
\affiliation{Department of Physics and Astronomy, Aarhus University, Ny
	Munkegade 120, 8000 Aarhus C, Denmark}
\author{Thomas Vosegaard}
\email[]{tv@chem.au.dk}
\affiliation{Interdisciplinary Nanoscience Center and Department of Chemistry, Aarhus University, Gustav Wieds Vej 14, 8000 Aarhus C, Denmark}
\date{\today}

\begin{abstract}
    We discuss the commonly encountered problem when optimizing NMR pulses using optimal control that the otherwise very precise NMR theory does not provide as excellent agreement with experiments. We hypothesize that this disagreement is due to phase transients in the pulse due to abrupt phase- and amplitude changes resulting in a large bandwidth. We apply the GROUP algorithm that gives high fidelity pulses with a low bandwidth compared to the typical GRAPE pulses. Our results obtain a better agreement between experiment and simulations supporting our hypothesis and solution to the problem.
\end{abstract}

\maketitle

\section{Introduction}
Nuclear Magnetic Resonance (NMR) is a routine tool in numerous disciplines ranging from materials sciences to food science and structural biology. While spin-1/2 nuclei like $^1$H, $^{13}$C, and $^{15}$N are most easily accessible, considerable information may be achieved by studying atoms with nuclear spin-1 or higher. These so-called quadrupolar nuclei experience a quadrupole coupling, which reports on the electric field gradients at the position of the nucleus. Quadrupolar nuclei constitute roughly 70\% of the periodic table with notable examples being spin-3/2 nuclei such as $^{23}$Na and $^{87}$Rb or the spin-5/2 nuclei $^{17}$O and $^{27}$Al. 

The major spectroscopic challenge with quadrupolar nuclei is the large quadrupole coupling leading to considerable broadening of the NMR lines, which in many cases cannot be removed by magic-angle spinning (MAS) \cite{Andrew_Nature_1958} due to the second-order effects. This challenge puzzled researchers for decades, but in the 90'es several solutions such as Dynamical-Angle Spinning \cite{Chmelka,Baltisberger} and Double Rotation \cite{Chmelka,Kentgens} were developed and they yield isotropic spectra for half-integer spin quadrupolar nuclei. However these methods are technically very demanding as they both involve a time-dependent sample-spinning axis. Alternatives are Multiple Quantum Magic-Angle Spinning (MQMAS) \cite{Frydman_2} and Satellite Transition Magic-Angle Spinning \cite{Gan_JACS_2000} that provide an isotropic dimension of a two-dimensional spectrum using conventional MAS.

The focus of this paper is the popular MQMAS, which is easy to implement and has high stability on many different samples. The only drawback of this experiment is its low sensitivity caused by difficulties in exciting the multiple-quantum coherence and the subsequent conversion into observable single-quantum coherence. The sensitivity issue has been the concern of many studies focusing on the excitation \cite{doi:10.1021/ja9614676, kwak_enhanced_2001, vosegaard_multiple_2001, Vosegaard} and mixing \cite{doi:10.1021/ja9614676, madhu_sensitivity_1999, vosegaard_sensitivity_2000, gan_enhancing_2004, kanwal_sensitivity_2019}. The 3Q $\rightarrow$ 1Q mixing is ideally performed using a selective inversion of the satellite transitions, which was the inspiration for the mixing sequence using fast amplitude modulated (FAM) pulses \cite{madhu_sensitivity_1999}. Excitation of the 3Q coherence is challenging, and even the best sequences generally yield efficiencies below 50\% leaving plenty of room for improvement. 

Quantum optimal control (OC) \cite{Glaser2015} has been used extensively in NMR \cite{doi:10.1002/9780470034590.emrstm1043} to improve sensitivity \cite{doi:10.1021/ja048786e, Vosegaard, doi:10.1002/ange.201805002} or to achieve specific features of the NMR experiment \cite{SKINNER20038, KEHLET2005204, KALLIES2018115}. The prevalent numerical OC algorithms used in NMR are the Gradient Ascent Pulse Engineering (GRAPE) \cite{GRAPE, doi:10.1063/1.4949534, properGRAPE}, Krotov \cite{krotov1995global, doi:10.1063/1.2903458}, and Chopped Random Basis (CRAB) \cite{CRAB}.  

Usually the 3Q coherence of the radio-frequency pulse is optimized using OC by dividing it into small steps and allowing the OC algorithm to vary the amplitude and phase of the rf-pulse steps. Typically, many OC optimizations are performed with random starting values for the controls to search the largest possible space and avoid local traps. Hence, we will often be equipped with a number of OC-pulses that perform well from a theoretical idealized point of view.

It is well-established that when implementing optimal-control NMR pulses experimentally, they often do not perform as well in experiment as in simulation, suggesting that the otherwise very precise theory describing the NMR spin-dynamics does not capture all details of the experiment. One feature lacking in the NMR theory typically used to calculate OC-pulse shapes is phase transients \cite{BARBARA1991497}, which have been shown to have significant impact in EPR \cite{SPINDLER201249}. These are commonly ignored in theoretical descriptions of NMR experiments, except a few cases \cite{doi:10.1021/ja016027f, VEGA200422, Emsley, WEBER201242, Ernst}. One of the difficulties in including phase transients in the theoretical description of NMR experiments is that they depend strongly on the hardware configuration and hence do not have a generic expression. Consequently, it seems like a better strategy to design experiments that avoid or reduce phase transients \cite{Ernst}. OC NMR experiments are expected to be particularly sensitive to phase transients, since they normally rely on short pulses of different phase and amplitude.

In this paper we hypothesise that phase transients may be problematic for the experimental implementation of OC-pulses, and that the effects of phase transients can be reduced by creating smooth OC-pulses instead of pulse shapes with many abrupt variations in either phase or amplitude. With inspiration in our previous work on improving the MQMAS efficiency by OC-pulse excitation \cite{Vosegaard}, we will test this hypothesis and present improved OC-pulse schemes. While our previous work employed the GRAPE algorithm, here we apply the GROUP algorithm \cite{GROUP} that introduces an optimization scheme contained in a reduced basis leading to smooth pulse shapes. This reduced basis may be any smooth basis, with natural choices being the CRAB Fourier basis \cite{CRAB, dCRAB}, Hermite polynomials \cite{motzoi2013, theis2018}, and Hanning windows  \cite{Theis2016}. It should be noted that the Spinach simulation software package allows to perform OC optimizations using different basis functions, although the specific possibilities are not documented \cite{Spinach2011}.

Here, we have the specific goal to limit the effect of phase transients that are presumably more pronounced for high-frequency components of the OC-pulse. To allow the best possible control of the frequency band, we have chosen to use a specific Fourier basis in the GROUP implementation, since this provides smooth pulses, enables control of the bandwidth and thereby limit phase transients resulting from the OC-pulse. Using GROUP, we improve the 3Q excitation efficiencies by about 50\% for $^{87}$Rb in the model compound RbClO$_4$ compared to traditional GRAPE methods. Furthermore, we show a clearer correlation between the theoretical and experimental excitation efficiencies with GROUP than using GRAPE. This indicates that phase transients are indeed a limiting factor in MQMAS experimentation. Our results suggest that finding high quality pulses with low bandwidth could be a principally straightforward way to improve experimental results.

\section{Theory \label{sec:Theory}}
\subsection{Hamiltonian and Objective}
We consider a basic Hamiltonian for a single nucleus spin $I>1/2$ in a solid under magic-angle spinning expressed in the Zeeman interaction frame
\begin{equation}
\hat{H}(t) = \hat{H}_{\text{iso}}^\sigma + \hat{H}_{\text{quad}}(t)+\hat{H}_{\text{rf}}(t).\label{eq:hamil}
\end{equation}
where the first two terms represent the isotropic chemical shift and quadrupole coupling Hamiltonians, respectively \cite{mehring2012principles}. The controllable part of the Hamiltonian is given by radio-frequency fields in the last term
\begin{equation}
\hat{H}_{\text{rf}}(t) = \omega_{\text{rf}}^x(t) \hat{I}_x + \omega_{\text{rf}}^y(t)  \hat{I}_y
\label{eq:H_rf}
\end{equation}
where $\hat{I}_{q}$ are spin angular momentum operators and $\omega^q_{\text{rf}}=\gamma B_\text{rf}^q$ are the components of the radio-frequency modulated magnetic field. 

The Hamiltonian for the isotropic chemical shift Hamiltonian is given by
\begin{equation}
\hat{H}_{\text{iso}}^{\sigma} = \omega_0 \delta_{\text{iso}} \hat{I}_z
\end{equation}
where $\omega_0 = \gamma_{\sigma}B_0$ is the Larmor frequency and $\delta_{\text{iso}}$ is the chemical shift. The Magnus expansion of the quadrupolar Hamiltonian after the transformation into the Zeeman interaction frame gives rise to the first- and second-order quadrupolar terms, where off-diagonal spin terms may be disregarded (secular approximation). The result is
\begin{equation}
\hat{H}_{\text{quad}}(\Omega_{\text{PR}}, \omega_r t) = \hat{H}_Q^{(1)}(\Omega_{\text{PR}}, \omega_r t) + \hat{H}_Q^{(2)}(\Omega_{\text{PR}}, \omega_r t),
\end{equation}
where $\Omega_{PR}=(\alpha_{PR}, \beta_{PR}, \gamma_{PR})$ is the set of Euler angles describing the orientation of the principal axis frame (P) of the quadrupole coupling tensor in the frame of the MAS rotor (R). Given that we consider a powder sample, the total transfer is represented by an integral over all crystallite orientations. In this work, we use sets of Euler angles, which have been obtained using the REPULSION method \cite{bak1997repulsion}. The first- and second-order terms of the quadrupolar Hamiltonian are given by
\begin{align}
\hat{H}_Q^{(1)} &= \omega_Q^{(1)}(t) \Bigl(3\hat{I}_z^2-I(I+1)\Bigr), \\
\hat{H}_Q^{(2)} &= \omega_Q^{(21)}(t) \Bigl(-8\hat{I}_z^2+4I^2-1\Bigr)\hat{I}_z \nonumber \\
&+ \omega_Q^{(22)}(t)\Bigl(-2\hat{I}_z^2+2I^2-1\Bigr)\hat{I}_z,
\end{align}
where expressions for the time-dependent coefficients $\omega_Q^{(i)}(t)$ are given in Ref.~\cite{SKIBSTED199188}.

The initial state is given by a Boltzmann distribution $\rho(0)=\mathbb{1}/4 + \hbar \omega_0/(kT) \hat{I}_z$. However, we can neglect the coefficients noting that the unit operator does not affect our NMR measurements and we always compare results for the same type of nuclei. The density operator at thermal equilibrium is represented by $\rho(0)=\hat{I}_z$. Focusing on excitation of 3Q coherence or efficiency, our goal is to maximize the 3Q coherence by maximising the projection $\mathcal{F}$ of the density operator ($\rho(T)$) onto the 3Q coherence operator ($\rho_{_t}$),
\begin{equation}
    \mathcal{F}=\frac{1}{\mathcal{N}}\bigl|\text{Tr}[\rho_t^\dagger \rho(T)] \bigr|^2 \label{eq:f}
\end{equation}
where $\mathcal{N}$ is a normalization factor such that $0 \le \mathcal{F} \le 1$. The target state given by
\begin{equation}
\rho_{_t} =  \begin{pmatrix}
0 & 0 & 0 & 1 \\
0 & 0 & 0 & 0 \\
0 & 0 & 0 & 0 \\
0 & 0 & 0 & 0
\end{pmatrix}.
\end{equation}

The objective in this paper is to design smooth pulse sequences $B_{\text{rf}}^x=\gamma\omega_{\text{rf}}^x(t)$ and $B_{\text{rf}}^y=\gamma\omega_{\text{rf}}^y(t)$ that maximize the projection onto the 3Q coherence state for the entire set of crystallites, which is a task well suited for optimal control theory.

\subsection{Parameter Robustness}\label{sec:Robustness}
We need pulses that are robust over a range of Hamiltonian parameters. In order to address this concern, an experimentally robust pulse is found by optimizing the expected coherence
\begin{equation}
\mathbb{E}[\mathcal{F}]=\sum_l p(\Omega_l) \mathcal{F}(\Omega_l) \label{eq:expectF}
\end{equation}
where $\Omega_l$ and $p(\Omega_l)$ are the members and probabilities of an appropriately chosen ensemble of Hamiltonians. In this section, we discuss how this ensemble is constructed. 

We discuss powder samples, i.e.~samples with all possible crystallite orientations present represented by $\Omega_\text{PR} = (\alpha_\text{PR}, \beta_\text{PR}, \gamma_\text{PR})$. The pulse sequence should be able to transfer the nuclear spin system from the initial to target state for all crystallites. In addition, rf-field inhomogeneity is a commonly encountered problem with all solid-state NMR probes and different samples display different nuclear spin interaction parameters. 

In the powder average we select $(\alpha_{\text{PR}},\beta_{\text{PR}})$ pairs using the the REPULSION algorithm from Ref.~\cite{bak1997repulsion}, while $\gamma_{\text{PR}}$  is selected in equidistant steps. Radio-frequency field inhomogeneity is taken into account by using an average of three rf-field strengths (95\%, 100\%, and 105\%). For this proof of concept, we have not made any efforts to make the sequences broadband with respect to the nuclear spin interaction parameters, but this would be straight forward to implement in a manner similar to the stability with respect to rf-inhomogeneity.

\subsection{Bandwidth Limitations}
The waveform generator of the NMR spectrometer generates shaped rf-pulses with a fine digitization, in our hardware down to 50 ns step sizes, which would allow to represent frequencies up to 1/50 ns = 20 MHz. This shaped pulse is then passed through amplifiers and filters into the NMR probe, which uses an LC circuit to transmit the power to the spins. The typical bandwidth of the probe is on the order of a few MHz implying that any high-frequency components of an rf-pulse may be truncated. Mehring and Waugh \cite{doi:10.1063/1.1685714} derived the response of an RCL circuit to the instant change in phase or amplitude of a pulse and demonstrated that such, instantaneous changes lead to phase transients, which later have been analysed theoretically in detail \cite{doi:10.1063/1.1680944,VEGA200422}. Since phase transients vary with the hardware configuration, the easiest way to circumvent problems arising therefrom is to create pulse shapes with no discontinuities and high-frequency components.

\subsection{Optimal Control Theory \label{sec:oct}}
In quantum optimal control, pulses that maximize the triple quantum coherence can be found by formulating the control program as a nonlinear optimization problem. In a general context, the goal is typically to choose a time-dependent control vector $\mathbf{u}(t)$  such that at the final time $t=T$ the final state $\rho(T)$ reaches the target state $\rho_t$ as closely as possible. In this context the controls are the rf-field amplitudes $\omega_\text{rf}^q(t)$, $q = x,y$. This condition can be quantified by the cost function
\begin{equation}
	J=1-\mathcal{F},\label{eq:overlap}
\end{equation}

An optimal control is found by iteratively updating the control such that the cost is minimized via
\begin{equation}
    \mathbf{u}^{(k+1)} = \mathbf{u}^{(k)} + \alpha^{(k)} \mathbf{p}^{(k)},
\end{equation}
where $\alpha^{(k)}$ is a properly chosen step size along the descent direction $\mathbf{p}^{(k)}$ for the \textit{k}th iteration. The descent direction is often based on gradient information where a simplistic choice would be the steepest descent direction $\mathbf{p}_k = - \nabla J(\mathbf{u}^{(k)})$. In practice, faster and better convergence is assured by quasi-Newton directions such as BFGS \cite{properGRAPE}. Multiple starting guesses $\mathbf{u}^{(0)}(t)$ are used (multi-starting), which for a large enough sample sizes give a decent probability that some of the optimized pulses come reasonably close to the global minimum in Eq.~\eqref{eq:overlap}. For an in-depth discussion of optimization we refer the reader to Ref.~\cite{nocedal2006numerical}.

Finding the controls that minimizes Eq.~\eqref{eq:overlap} is a classical numerical optimization with high quality implementation available in numerous software packages. The outstanding challenge is to reliably, efficiently and  accurately compute the gradient ($\nabla J(\mathbf{u}_k)$) that is given to the optimization algorithm. The standard methodology for computing such gradients is GRAPE \cite{GRAPE,properGRAPE}, which we summarize here, as well as enhancements for bandwidth reduction \cite{GROUP, MotzoiGROUP}. We focus on the very common scenario where the underlying Hamiltonian has a bilinear dependence on the control
\begin{equation}
	\hat{H}(t) = \hat{H}_0 + \sum_k u_k(t) \hat{H}_k.\label{eq:hamilt}
\end{equation} 
where $\hat{H}_0$ is the uncontrollable drift Hamiltonian and $\hat{H}_k$ are the controllable Hamiltonians. The standard NMR Hamiltonian used above \eqref{eq:hamil} can exactly be cast in this form.

In order to minimize Eq.~\eqref{eq:overlap} we must solve the Liouville-von-Neumann equation for the dynamics $\dot{\rho}=-i[\hat{H},\rho]$. Generally, this equation has no analytic solutions when $\hat{H}_0$ is time-dependent. Numerical solutions rely on some discretization of time. This presents us with two different choices, i) we first discretize the dynamics and then find expressions for the gradients as in Refs.~\cite{GRAPE,properGRAPE} or ii) we find expressions for the gradients that we then discretize as in Refs.~\cite{qEngine, GOAT}. The former approach is more exact and can benefit in convergence while the latter is faster to compute, independent of dicretization, and works especially well in the small $\Delta t$ limit. In this paper we will follow the first approach.

We assume the control is piecewise constant over small interval of length $\Delta t$. This amounts to performing the replacement
\begin{equation}
u_{k}(t) \leftarrow \sum_{j=0}^{N-1} u_{k,j}\sqcap_j(t,\Delta t), \label{eq:discretization}
\end{equation}
with the rectangle function
\begin{equation}
	\sqcap_j(t,\Delta t)\equiv\left[\Theta(t-j \Delta t) - \Theta(t-(j+1) \Delta t) \right],
\end{equation}
where $\Theta$ is the Heaviside unit step function. The solution to the Liouville-von-Neumann equation over a short time-interval is
\begin{equation}
    \rho(t_j+\Delta t) = \hat{U}_j \rho(t_j) \hat{U}_j^\dagger,
\end{equation}
where $\hat{U}_j = \exp\big(-i (\hat{H}_0(t) + \sum_k u_{k,j}\hat H_k) \Delta t\big)$. The final state can be found straightforwardly as
\begin{equation}
    \rho(T) = \hat{U}(T) \rho(0) \hat{U}^\dagger(T),
\end{equation}
with $\hat{U}(T) = \Pi_{j=0}^{N-1} \hat{U}_n = \hat{U}_{N-1}\hat{U}_{N-2} ... \hat{U}_1 \hat{U}_0$.

Having discussed the temporal discretization we proceed to deriving the gradient of Eq.~\eqref{eq:hamilt} with respect to $u_{k,j}$. Utilizing the cyclic property of the trace and the temporal discretization allows us to rewrite,
\begin{equation}
    J = 1 - \frac{1}{\mathcal{N}}\Bigl|\text{Tr}[\rho_t^\dagger(t_{j+1}) \hat{U}_j \rho(t_{j}) \hat{U}_j^\dagger]\Bigr|^2,
\end{equation}
where
\begin{align}
    \rho(t_j) &= \Biggl(\prod_{n=0}^{j-1}\hat{U}_n\Biggr) \rho(t_0) \Biggl(\prod_{n=j-1}^{0}\hat{U}_n^\dagger\Biggr), \\
    \rho_t^\dagger (t_j) &= \Biggl(\prod_{n=N-1}^{j}\hat{U}_n^\dagger\Biggr) \rho_t^\dagger(t_N) \Biggl(\prod_{n=j}^{N-1}\hat{U}_n\Biggr),
\end{align}
where $\rho^\dagger(t_N)=\rho_t^\dagger$ and $\rho(t_0)=\rho(0)$. These are respectively the forward propagated initial state and the backwards propagated target state. A short calculation gives 
\begin{align}
    \frac{\partial J}{\partial u_{k,j}} =-\frac{2}{\mathcal{N}} \textrm{Re}&\biggl( \mathrm{Tr}\biggl[\rho_{t}^\dagger(t_{j+1})\biggl(\frac{\partial \hat{U}_j}{\partial u_{k,j}} \rho(t_{j}) \hat{U}_j^\dagger \nonumber \\
&+ \hat{U}_j \rho(t_{j}) \frac{\partial \hat{U}_j^\dagger}{\partial u_{k,j}} \biggr)\biggr] \text{Tr}\Bigl[\rho_{t}\rho(T)^\dagger\Bigr]\biggr), \label{eq:gradient}
\end{align}

A convenient property of GRAPE is that all density operators $\rho(t_j)$ and $\rho_t^\dagger(t_j)$ can be updated by a single forwards and backwards propagation. The full gradient is then found by evaluating Eq.~\eqref{eq:gradient} at all points in time i.e. for each \textit{j}. This permits a large speed up compared to naive approach where one queries the cost function for each variation in the control values that would require at least \textit{N} propagations. Evaluating Eq.~\eqref{eq:gradient} also requires an expression for the derivative $\partial \hat{U}_j/\partial u_{k,j}$. Here we use the exact gradient from Ref.~\cite{properGRAPE}, being important for fast quasi-Newton methods. The full calculation is given in the appendix and the result is
\begin{align}
    \frac{\partial \hat{U}_j}{\partial u_{k,j}} &= \hat{U}_j \hat{V}_j \Bigl(\bigl(\hat{V}_j^\dagger \bigl(-i \hat{H}_k \Delta t\bigr) \hat{V}_j\bigr) \odot \hat{G}_j \Bigr) \hat{V}_j^\dagger \nonumber \\
    &= \hat{U}_j \hat{D}_j \label{eq:derivative}
\end{align}where $\hat{V}_j = \sum_l |\phi_l\rangle \langle l|$ is a unitary transformation diagonalizing the instantaneous Hamiltonian with eigenstates $\hat{H}(t_j)|\phi_l\rangle = E_j^l |\phi_l\rangle$ in a reference basis $\{|l\rangle\}$. The element-wise Hadamard product is $\odot$ and $\hat{G}_j$ is a matrix with entries
\begin{equation}
\Braket{m|\hat{G}_j|n}=
\begin{cases}
    1 & \text{for } E_j^m = E_j^n, \\
\frac{\exp(i(E_j^m-E_j^n)\Delta t)-1}{i(E_j^m-E_j^n)\Delta t} & \text{otherwise}.
\end{cases}
\end{equation}
Plugging Eq.~\eqref{eq:derivative} into Eq.~\eqref{eq:gradient} and rearranging gives the final form of the exact gradient
\begin{align}
    \frac{\partial J}{\partial u_{k,j}} =-\frac{2}{\mathcal{N}} \textrm{Re}&\biggl( \mathrm{Tr}\biggl[\rho_{t}^\dagger(t_{j})\biggl(\hat{D}_j \rho(t_{j})
    + \rho(t_{j}) \hat{D}_j^\dagger\biggr)\biggr] \nonumber \\
    &\times \text{Tr}\Bigl[\rho_{t}\rho(T)^\dagger\Bigr]\biggr) \label{eq:final:derivative}
\end{align}

In GRAPE we optimize the value of the control in each small piecewise constant interval, which often leads to controls with large variations between adjacent intervals, i.e.~$u_{k,j}$ and $u_{k,j+1}$, as there is no regularization limiting such variations in the standard formulation of GRAPE. 

To alleviate this problem, we will perform the optimization in Fourier space with a reduced basis using the GROUP methodology \cite{GROUP}, which as shown below produces controls without high frequency components, enabling us to sidestep the detrimental phase transients.

In GROUP we keep the discretization from Eq.~\eqref{eq:discretization} but parametrize the piecewise constant values as
\begin{equation}
    u_{k,j} = \sum_{m=0}^M c_{k,m} f_{k,m}(j\Delta t) S_k(j\Delta t),
\end{equation}
where $f_{k,m}(t)$ are a set of $M$ preferably smooth basis functions and $S_k(t)$ is a shape function zeroing the control at the boundaries of the time interval if needed. The central point is that we perform the optimization with respect to $c_{k,m}$ rather than $u_{k,j}$. If the basis size \textit{M} is sufficiently small with well chosen basis functions then the resulting control also becomes smooth. As in GRAPE we must provide the optimization algorithm with accurate analytic gradients and the derivative is straightforwardly found \cite{MotzoiGROUP} using the chain rule 
\begin{equation}
    \frac{\partial J}{\partial c_{k,m}} = \sum_{j=0}^{N-1} \frac{\partial J}{\partial u_{k,j}} \frac{\partial u_{k,j}}{\partial c_{k,m}}
    = \sum_{j=0}^{N-1} \frac{\partial J}{\partial u_{k,j}} f_{k,m}(j\Delta t) S_k(j\Delta t),
    \label{eq:GROUP}
\end{equation}
which is reminiscent of a discretized time-ordered integral. In GROUP, we then start with computing the regular GRAPE gradient in Eq.~\eqref{eq:final:derivative} \cite{GROUP}, apply the chain rule \eqref{eq:GROUP}, and perform the optimization iteration over $c_{k,m}$. The partial derivative $\partial u_{k,j}/ \partial c_{k,m}=T_{k,m,j}$ is referred to as the response function. If the exact linear response function is known, e.g. the actual experimental low-pass filter, this function can be substituted here \cite{MotzoiGROUP}.

In the following optimization, we choose the basis functions as Fourier components for each \textit{k}
\begin{equation}
f_{k,m}(t) =  \sin\biggl(\frac{(m+1)\pi  t}{T}\biggr). \label{cont_fourier_transform}
\end{equation}
These are always zero at the boundary of the time interval so we simply set $S_k(t)=1$.

As discussed above there is intrinsic parameter uncertainties in solid-state NMR experiments due to e.g. different crystal orientations and isotropic shifts. In particular, we need to optimize the expected coherence $\mathbb{E}[\mathcal{F}]$. To account for this, we define the expected cost function
\begin{equation}\label{eq:robustphi}
    \mathbb{E}[J]=1-\mathbb{E}[\mathcal{F}] = \sum_l p(\Omega_l) J(\Omega_l),
\end{equation}
where $J(\Omega_l)$ is the cost of a single member in the ensemble. This is a combination of all the possible $J(\Omega_j)$ so the gradient of $\mathbb{E}[J]$ is
\begin{equation}
\frac{\partial \mathbb{E}[J]}{\partial u_{k,j}}  = \sum_l p(\Omega_l) \frac{\partial J({\Omega_l})}{\partial u_{k,j}}.
\end{equation} 
The associated GROUP gradient of $\mathbb{E}[J]$ is found by replacing \textit{J} with $\mathbb{E}[J]$ in Eq.~\eqref{eq:GROUP}.

\section{Experimental methods\label{sec:experimental}}
All experiments have been carried out with a 400 MHz spectrometer with a Larmor frequency for $^{87}$Rb at $\omega_0/2\pi =$ 130.9 MHz. The sample used is RbClO$_4$, purchased from Sigma-Aldrich and used without further purification, which has well-characterized nuclear spin interaction parameters \cite{vosegaard_combined_1995, vosegaard_quadrupole_1996}. 

Chemical shifts are referenced to 1M RbNO$_3$ at 0 ppm. Calibrations of the power level has been carried out by measuring the 90 and 180 degree flip angles in the 1M RbNO$_3$ sample and for low powers on the solid RbClO$_4$ sample, where the double nutation frequency is obtained. These two measurements, however, do not provide very precise values for the rf-field strength, in the first case because the Q factor of the probe changes between the liquid and solid sample, and for the latter approach since it is necessary to work in the low-power regime. Fine-tuning of the rf-field strength was achieved by scanning through the FASTER condition \cite{FASTER}. In particular, finding the resonance conditions with zero intensity for rf-field strengths equal to half-integer multiples of the spin rate turns out to provide very accurate relations between pulse power and rf-field strength.

One-dimensional triple-quantum filtered MAS NMR experiments (corresponding to the MQMAS NMR experiment with $t_1 = 0$) were performed in a 2.5mm rotor spinning 30 kHz. The pulse sequence is based on the shifted-echo sequence \cite{Vosegaard2} with a shaped pulse for the 3Q coherence excitation followed by the 3Q to 1Q mixing using the fast amplitude modulation (FAM) sequence \cite{FAM} and subsequently an echo with 5 ms delay to ensure truncation-free whole-echo acquisition. The FAM mixing was achieved with 3 cycles of $x-\tau-\overline{x}-\tau$ with pulses and delays of $0.75\,\mu$s and 220 kHz rf-field strength, which we find to give the best possible $3Q\rightarrow 1Q$ mixing. All experiments employed 48 scans, with a $0.5$s repetition delay and were recorded at room temperature.

\begin{figure}[t]
\centering
\includegraphics[width=0.45\textwidth]{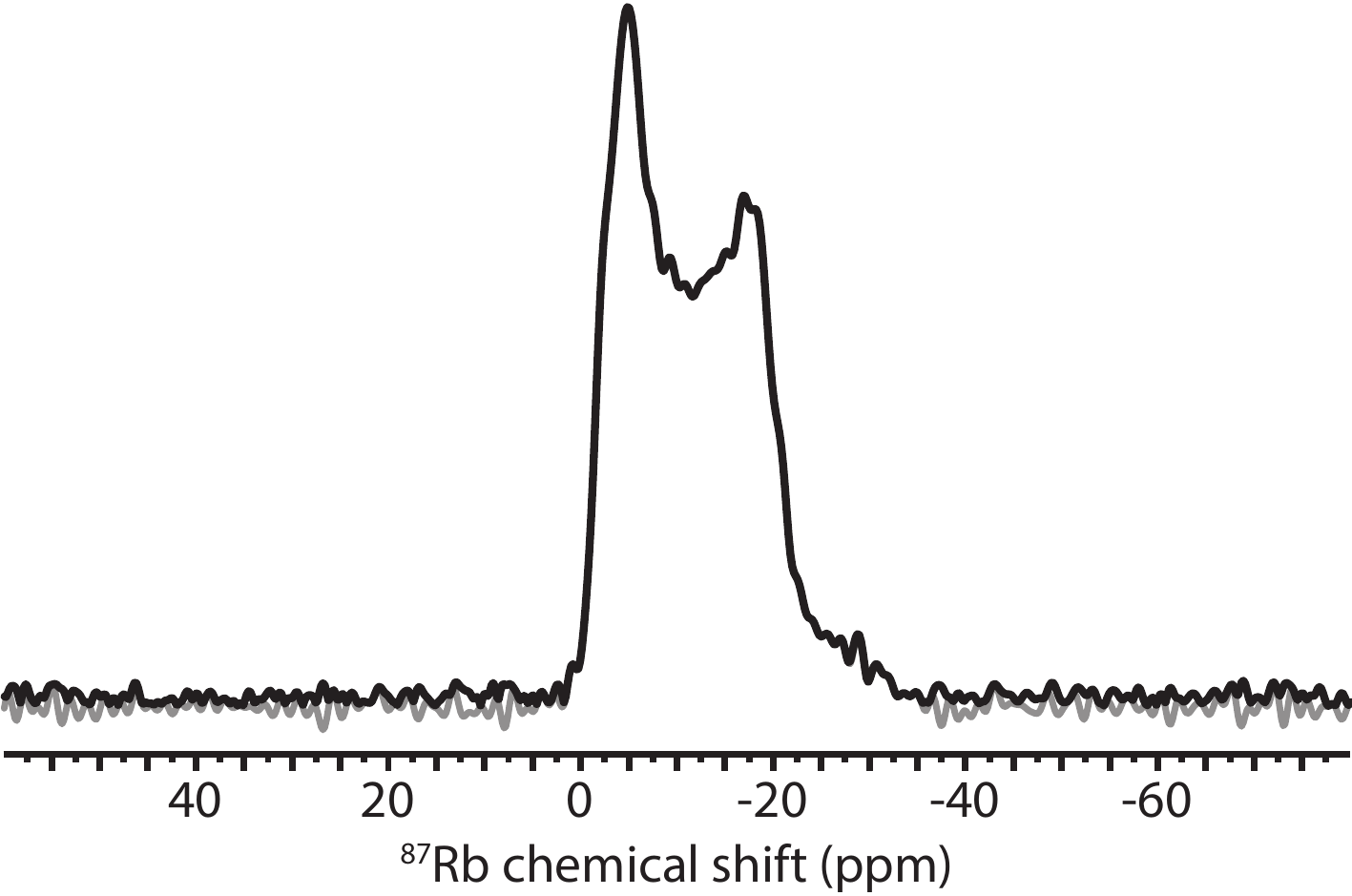}
\caption{Experimental $^{87}$Rb NMR spectrum of RbClO$_4$ recorded with a spin-echo pulse sequence using an echo time of 5 ms from whole-echo acquisition. The gray spectrum is obtained by phasing the spectrum to pure absorption mode, and the black spectrum is obtained by representing the complex spectrum in magnitude mode.}
\label{fig:spec}
\end{figure}

\begin{figure*}[t]
    \begin{minipage}[t]{0.49\textwidth}
       \includegraphics[scale=1.0]{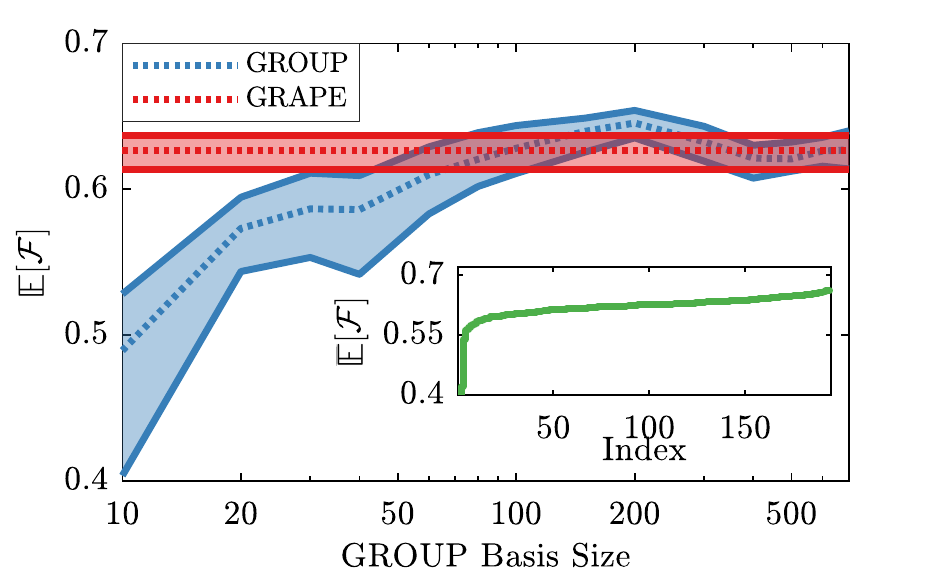}
       \caption{(color online)  Simulated efficiencies of optimized pulses using a truncated Fourier basis.  The blue dotted line shows the median of the GROUP solutions (of 100 individual optimizations). The GROUP solutions improve with increasing bandwidth. The red solutions (dotted line) are those given by GRAPE, which do not have bandwidth limitations under a piecewise constant approximation but are shown for comparison. In both cases, the shaded area represents the interquartile (25\%, 50\%, and 75\%) range of the 100 optimizations for each basis size. The range of GRAPE solutions is also plotted individually in green in the inset.}
       \label{fig:results1}
    \end{minipage}
    \;
    \begin{minipage}[t]{0.49\textwidth}
        \centering
        \includegraphics[scale=1.0]{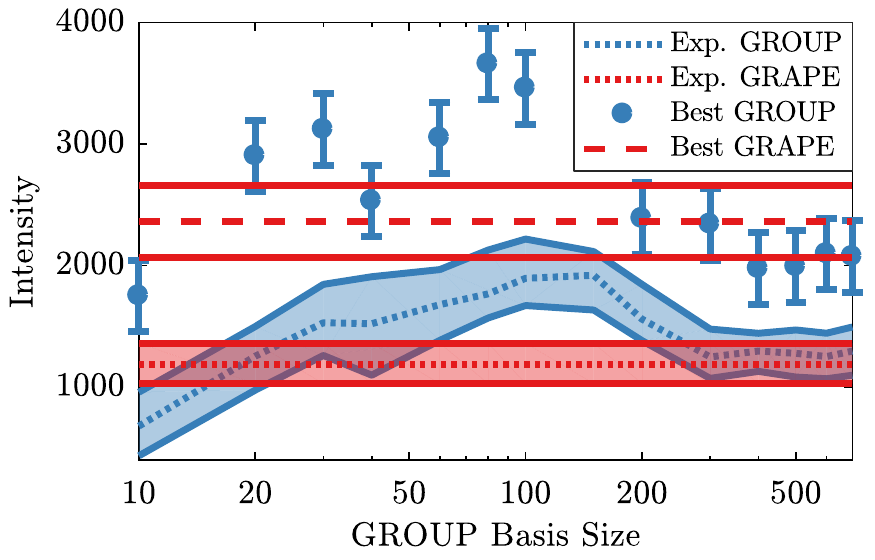}
        \caption{(color online)  Comparison of experimental results for GROUP and GRAPE pulses. The colour scheme for the dotted lines and shaded areas are the same as in Fig. \ref{fig:results1}.  The best experimental output is also plotted for each basis size for GROUP (out of 100), while the horizontal red line shows the best GRAPE pulse (out of 200). As with the theory predictions, GRAPE does not contain bandwidth limitations and thus shows no variation with basis size.}
        \label{fig:results2}
    \end{minipage}
\end{figure*}

Fig.~\ref{fig:spec} shows an experimental $^{87}$Rb NMR spectrum of RbClO$_4$ recorded using a conventional spin-echo sequence and whole-echo acquisition and illustrates the second-order quadrupolar lineshape. In the following the experimental spectra are processed by Fourier transform and represented as magnitude mode. The total intensity is then calculated as the integral of the signal in the range $-60$ ppm to $0$ ppm by subtracting the DC offset resulting from the magnitude calculation. The DC offset is calculated as the average intensity in the range 20 ppm to 80 ppm.

\section{Results and Discussion}
\subsection{Theoretical optimizations}
\begin{figure}[t]
\centering
\includegraphics[width=0.45\textwidth]{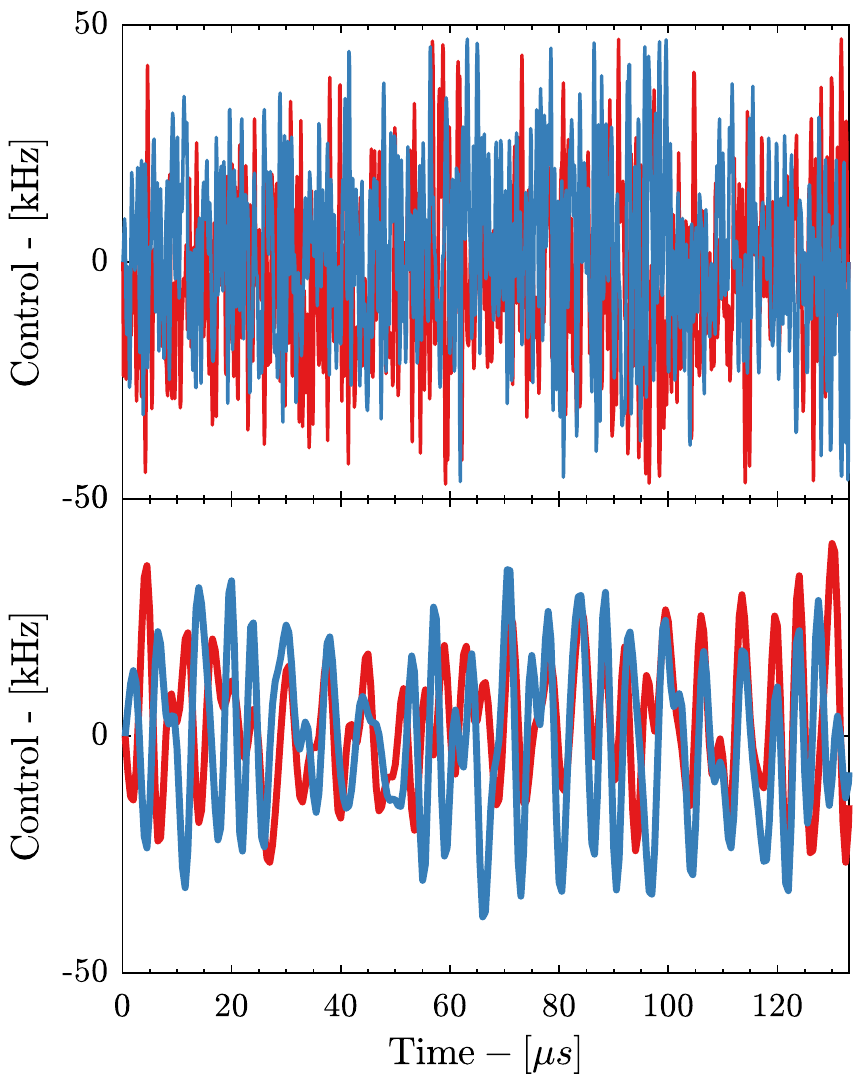}
\caption{(color online)  Optimal pulse for GRAPE (top) and GROUP (bottom). The blue line is $\omega_{\text{rf}}^x$ while the red is $\omega_{\text{rf}}^y$. The GROUP pulse is for a basis size of 100.}
\label{fig:resultsControl}
\end{figure}

\begin{figure}[t]
\centering
\includegraphics[scale=1.0]{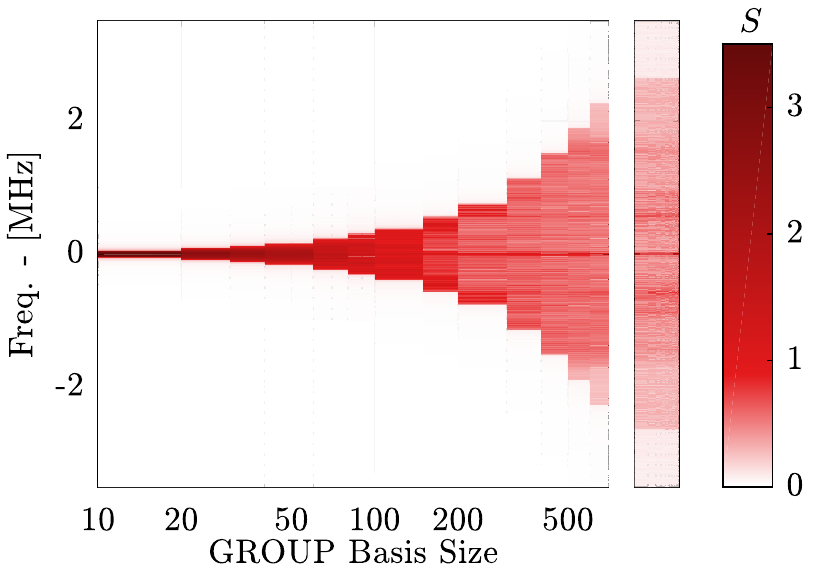}
\caption{(color online) Average absolute value of the Fourier transformed pulses as a function of basis size for the two controls $S=\mathbb{E}[|\text{FT}(\omega_{rf}^x+i \omega_{rf}^y)|]$ where $\text{FT}$ denotes the Fourier transformation. The left plot is for GROUP as a function of basis size, while the middle bar graph is for GRAPE. The spectra are averaged over all optimized pulses for each basis.}
\label{fig:fourier}
\end{figure}

Our first step in creating a new OC-pulse for improved 3Q excitation in MQMAS experiments is to perform the numerical optimization using the GROUP algorithm outlined above. The OC-pulses are optimized for a quadrupole coupling tensor with $C_Q = 3.2$ MHz and $\eta_Q = 0.2$ corresponding to the parameters for RbClO$_4$. To ensure stability towards hardware imperfections, we apply rf-field inhomogeneity as described earlier. Our strategy has been to derive a large number of pulses resulting from optimizations employing different random starting values for the controls to ensure good coverage of the parameter space. For GROUP, we have performed 100 optimizations for each basis size. For comparison, we have performed 200 GRAPE optimizations employing 1331 piecewise constant controls with a step $\Delta t = 0.1\mu$s. The results of these optimizations are summarized in Fig.~\ref{fig:results1} where the GROUP results are plotted as function of the Fourier basis size. The figure does not report results of the individual optimizations, rather it displays the quartiles of transfer efficiency for each basis set size. Note that GRAPE does not have a Fourier basis size, hence it is constant along the $x$-axis of the plot but it is shown for reference.

It is clear from this plot that GRAPE and GROUP converge on solutions with similar fidelities in the high frequency domain, while smaller bases severely limit the performance of the pulses indicating that there are not enough degrees of freedom to find good solutions to the problem. On the other hand, the maximum at high frequency is slightly larger for GROUP with a basis set size of 200, which we attribute to the smooth pulses being a slightly better ansatz than piecewise constant controls. The inset of Fig.~\ref{fig:results1} shows the distribution of fidelities in the GRAPE results, which is almost entirely in the 60\% range, indicating that it is most likely the best we can expect to do theoretically given the ensemble size.

\subsection{Experimental Results}
The experimental examination of the OC-pulses found is key to evaluate if our hypothesis that phase transients of the rf-pulses are responsible for the potential disagreement between the experimental and theoretical results for the OC-pulses. The experimental intensities for the 3Q excitation are evaluated for a total of 1400 GROUP pulses with 100 pulses for each basis set size and for 200 GRAPE pulses. Ideally, there should a linear correlation between the theoretical average coherence $\mathbb{E}[\mathcal{F}]$ and the experimental intensity.

\begin{figure}[t]
\centering
\includegraphics[width=0.45\textwidth]{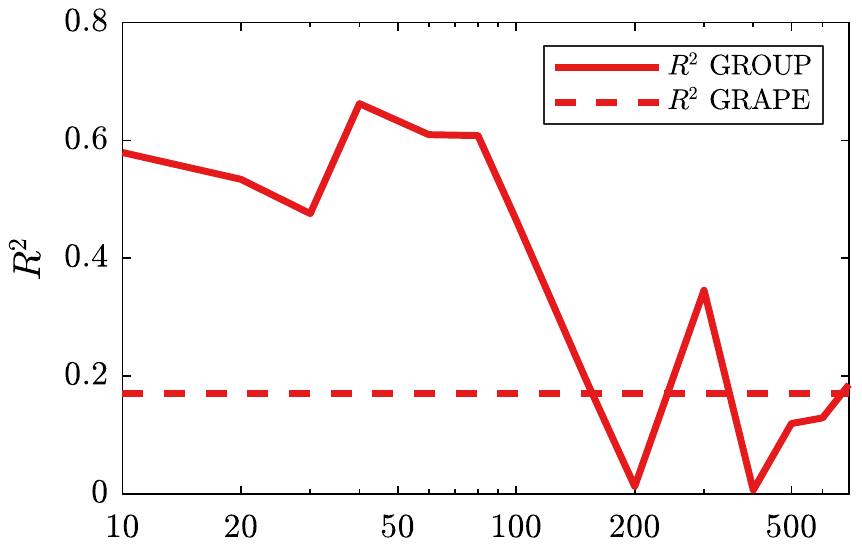}
\caption{(color online)  Comparison of experimental outcomes vs. theoretical predictions for sampled pulse sequences. The red solid and dashed lines plot the $R^2$ correlation for GROUP and GRAPE. Note the theoretical and experimental values correlate up to a cutoff basis size after which filtering effects greatly reduce the model validity.}
\label{fig:results3}
\end{figure}

Fig.~\ref{fig:results2} shows the experimentally obtained performance of the GRAPE and GROUP pulses. The best results are highlighted by individual points while the shaded area show the distributions as 25\% to 75\% quartiles. In both the best and average case, it is clear that GROUP performs best with basis size around 100.  Meanwhile GRAPE is below GROUP, from basis size 20 up until 200. At larger basis sizes we see that GROUP and GRAPE perform with similar efficiency, suggesting that high frequencies components of the GROUP pulses degrade their performance.

The measurement uncertainty is calculated for individual pulses based on experimental drift measured during the entire experimental run. This is estimated by running the 200 GRAPE pulses both before and after the much larger number of 1400 GROUP pulses (spread over 14 different basis sizes). The changes in GRAPE output intensity is binned and a Gaussian distribution is extracted. The standard deviation ($1\sigma$) of the resulting distribution is used for the error bars in  Fig.~\ref{fig:results2}.

The optimal GRAPE and GROUP pulses are plotted in Fig.~\ref{fig:resultsControl} (top) and (bottom) respectively. The GROUP pulse performs more than 50\% better than the GRAPE pulse. It is reasonable to expect, given the problems with phase transients and the general trends in Figs. \ref{fig:results1} and \ref{fig:results2}, that the higher bandwidth in the GRAPE pulses are the main reason for the lower performance.

To investigate the possible (mis)performance of the different pulses, suspecting that truncation of high-frequency components is a main source, we have plotted the frequency-distribution of the different pulses (average absolute spectrum on the colour bar) of the GROUP and GRAPE rf-pulses in Fig. \ref{fig:fourier}. In the GROUP case, we see a clear increase in bandwidth with basis size (left plot), as expected, while the GRAPE average spectrum remains wider than even the largest basis size for GROUP (middle bar). These plots clearly demonstrate that it is indeed possible to design highly efficient pulses without the need for a high bandwidth, which would be truncated by the rf-circuit.

Finally, we validate that the time-dependent Hamiltonian with lower frequency components more closely matches the experimentally observed spin dynamics. To investigate this, we perform a linear fit between the numerical simulations and experiment and calculate the coefficient of determination $R^2$. While the experimental data is still quite noisy, we see that the low-frequency GROUP data provides reasonable agreement between the experimental and theoretical transfer efficiencies ($R^2=0.6$), which is significantly better than the GRAPE-fit ($R^2=0.18$). Moreover, the quality of the GROUP fit dramatically decreases for higher basis size. Therefore, we conclude that removing high frequencies from the rf-pulses results in significantly better agreement between the experiment and theoretical models.

\section{Conclusion \label{sec:conclusion}}
This work shows that increased precision may be obtained in optimal control NMR pulses by incorporating robustness to phase transients. First, we see a 50\% increase in efficiency of the presently proposed GROUP pulses compared to standard GRAPE techniques. Secondly, we see a stronger correlation between optimal control theory and experiments. Moreover, our hypothesis that higher bandwidths lead to performance degradation is corroborated with GRAPE pulses performing similarly to high-frequency GROUP pulses. The performance of the smooth pulses help to validate their use in high-precision NMR since they sidestep the nonlinear phase transients that are difficult to model. This also points the way to using similar techniques in other quantum systems for improved optimal control, especially where precise modeling of the phase transients is too demanding.

\section{Acknowledgements}
The authors would like to thank M. Dalgaard and J. H. M. Jensen for carefully proof reading the manuscript. We acknowledge funding from the Carlsberg, John Templeton foundations, ERC, H2020 grant 639560 (MECTRL), and the Danish Ministry of Higher Education and Science (AU-2010-612-181).

\appendix
\section{Exact Derivative}
In this appendix we review the calculation for the derivative of the time evolution operator $\hat{U}_j$ originally presented in Ref.~\cite{properGRAPE}.

It is non-trivial to compute this derivative since the Hamiltonian and its derivative need not commute. Expanding the propagator as a Taylor series and retaining the ordering of the operators yields
\begin{align}
	\frac{\partial \hat{U}_j}{\partial u_{k,j}} &= \sum_{n=1}^{\infty} \frac{ \left( -i \Delta t \right) ^n }{n!} \sum_{m=0}^{n-1} \hat{H}(t_j)^m \hat{H}_k \hat{H}(t_j)^{n-m-1} \nonumber \\
	&= \sum_{n=0}^{\infty} \sum_{m=0}^{\infty} \frac{A^n B A^m}{(n+m+1)!} \; , \label{eq:derivTaylorExp2}
\end{align} 
where $A = -i \hat{H}(t_j) \Delta t$ and $B = -i \hat{H}_k \Delta t$ have been defined for convenience. In order to continue we use the identity
\begin{equation}
	\int_{0}^{1} (1-\alpha)^n \alpha^m \mathrm{d}\alpha  = \frac{n! m !}{(n+m+1)!}
\end{equation}
Thereby Eq.~\eqref{eq:derivTaylorExp2} can be expressed as
\begin{align}
	\frac{\partial \hat{U}_j}{\partial u_{k,j}} &= \sum_{n=0}^{\infty} \sum_{m=0}^{\infty} \frac{A^n B A^m}{n! m!}  \int_{0}^{1} (1-\alpha)^n \alpha^m \mathrm{d}\alpha \nonumber \\
	&= \int_{0}^{1} \sum_{n=0}^{\infty} \sum_{m=0}^{\infty} \frac{(A (1- \alpha))^n}{n!} B \frac{(A \alpha)^m}{m!}  \mathrm{d}\alpha \nonumber \\
	&= e^A \int_{0}^{1} e^{ - \alpha A} B e^{ \alpha A} \mathrm{d}\alpha, \label{eq:derivTaylorExp3}
\end{align}
Although Eq.~\eqref{eq:derivTaylorExp3} is rather compact, evaluating the integral in its current form is complicated. Instead, the integral can be explicitly solved by applying the Baker-Campbell-Hausdorff expansion 
\begin{equation}
	e^X Y e^{-X} = \sum_{l = 0}^{\infty} \frac{ [ X,Y  ]_l }{l!} = Y + [ X,Y  ] + \frac{1}{2!} [ X , [ X,Y  ]] + ...
\end{equation}
where $[ X , Y ]_l = [ X ,[ X , Y]_{l-1}]$ and $[X,Y]_0 = Y$ is the definition of the recursive commutator \cite{boas2006mathematical}. Evaluating the integral Eq.~\eqref{eq:derivTaylorExp3} becomes
\begin{equation}
    \frac{\partial \hat{U}_j}{\partial u_{k,j}} = -i\Delta t \hat{U}_j  \sum_{l = 0}^{\infty}  \frac{i^{l} \Delta t^{l}}{(l+1)!} [\hat{H}(t_j),\hat{H}_k]_l. \label{eq:bakerExpansion}
\end{equation}
For sufficiently small time-steps the higher order corrections contributions can be neglected. However, decreasing the time-step increases the run-time of the time-evolution. Instead the traditional approach is to evaluate Eq.~\eqref{eq:derivTaylorExp3} by a change of basis into a Hamiltonian eigenbasis. The instantaneous eigenstates are $\hat{H}(t_j)|\phi_l\rangle = E_j^l |\phi_l\rangle$ and $\hat{V}_j^\dagger \hat{H}(t_j) \hat{V}_j = \hat{\Lambda}_j$ where $\hat{\Lambda}_j$ is a diagonal matrix and $\hat{V}_j = \sum_l |\phi_l\rangle \langle l|$ is a unitary transformation from the current basis $\{|l\rangle\}$ into the eigenbasis. Using $\mathbb{1}=\hat{V}_j \hat{V}_j^\dagger$ in Eq. \eqref{eq:derivTaylorExp3} gives
\begin{equation}
    \frac{\partial \hat{U}_j}{\partial u_{k,j}} = -i \Delta t \hat{U}_j \hat{V}_j \int_0^1 e^{\alpha i \hat{\Lambda}_j \Delta t} \hat{V}_j^\dagger \hat{H}_k \hat{V}_j e^{-\alpha i \hat{\Lambda}_j \Delta t} \text{d}\alpha \hat{V}_j^\dagger.
\end{equation}
Consider each of the matrix-elements of integral in the current basis
\begin{align}
    I_{m,n}&=\Braket{m|\int_0^1 e^{\alpha i \hat{\Lambda}_j \Delta t} \hat{V}_j^\dagger \hat{H}_k  \hat{V}_j e^{-\alpha i \hat{\Lambda}_j \Delta t} \text{d}\alpha|n} \nonumber \\
&=\int_0^1 \Braket{\phi_m|\hat{H}_k|\phi_n} e^{i \alpha(E_j^m-E_j^n)\Delta t}\text{d}\alpha \nonumber \\
&=
\begin{cases}
    \bigl\langle \phi_m\bigl|\hat{H}_k\bigr|\phi_n\bigr\rangle & \text{for } E_j^n=E_j^m, \\
    \bigl\langle \phi_m \bigl|\hat{H}_k\bigr|\phi_n\bigr\rangle \frac{\exp(i(E_j^m-E_j^n)\Delta t)-1}{i(E_j^m-E_j^n)\Delta t} &  \text{otherwise}.
\end{cases}
\end{align}
This allows us to rewrite the exact derivative as
\begin{equation}
    \frac{\partial \hat{U}_j}{\partial u_{k,j}} = \hat{U}_j \hat{V}_j \Bigl(\bigl(\hat{V}_j^\dagger \bigl(-i \hat{H}_k \Delta t\bigr) \hat{V}_j\bigr) \odot \hat{G}_j \Bigr) \hat{V}_j^\dagger.
\end{equation}
where $\odot$ is the element-wise Hadamard product and $\hat{G}_j$ is a matrix with entries
\begin{equation}
\Braket{m|\hat{G}_j|n}=
\begin{cases}
    1 & \text{for } E_j^n=E_j^m, \\
\frac{\exp(i(E_j^m-E_j^n)\Delta t)-1}{i(E_j^m-E_j^n)\Delta t} & \text{otherwise}.
\end{cases}
\end{equation}
Expanding the exponential to first order in $\Delta t$ gives $\langle m | \hat{G}_j |n\rangle = 1$ and the derivative becomes the same as only keeping the first term in \eqref{eq:bakerExpansion}.

\bibliography{lit_NMRGROUP}

\end{document}